\documentclass[a4paper,12pt]{article}
\usepackage{fullpage,epsf}
\usepackage{graphicx,caption}
\usepackage{bbold}
\usepackage{amsmath}
\usepackage{float}
\usepackage{graphicx}
\usepackage{mathtools}
\usepackage{physics}
\usepackage{url}
\usepackage{qcircuit}
\usepackage{multirow}
\usepackage{tabu}
\usepackage{listings}
\usepackage{lipsum}
\usepackage{subcaption}

\usepackage[backend=biber,style=phys,biblabel=brackets,sorting=none]{biblatex}
\usepackage{amsmath}

\begin{document}
\pagestyle{empty}                       
\epsfxsize=40mm                         

\begin{minipage}[b]{110mm}
        {\Huge\bf School of Physics\\ and Astronomy
        \vspace*{17mm}}
\end{minipage}
\hfill
\begin{minipage}[t]{40mm}               
        \makebox[40mm]{
        \includegraphics[width=40mm]{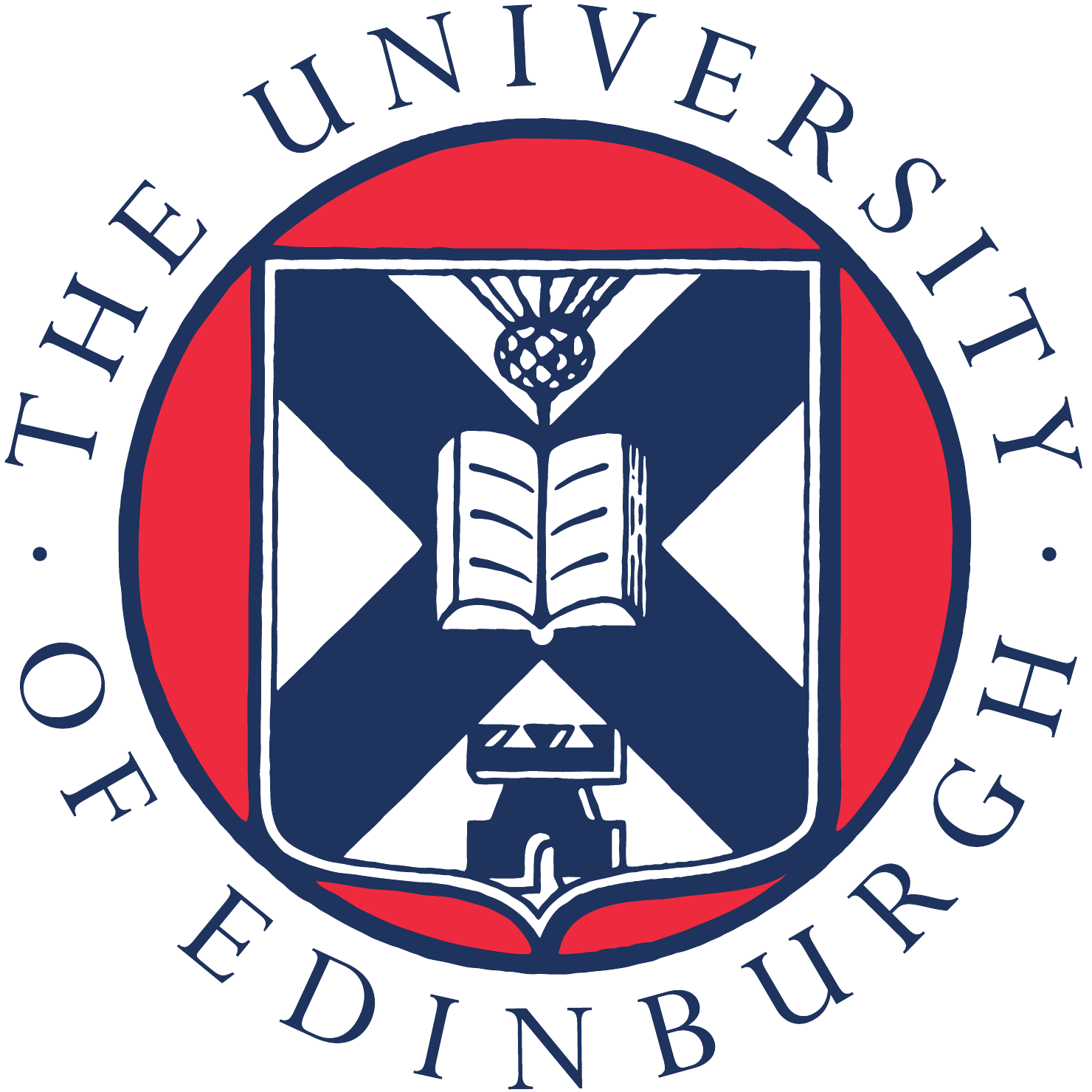}}
\end{minipage}
\par\noindent                                           
\begin{center}
        \Large\bf \Large\bf Senior Honours Project\\
        \Large\bf Computational Physics\\[10pt]                     
        \LARGE\bf Simulation of Networked Quantum Computing on Encrypted Data         
\end{center}
\vspace*{0.5cm}
\begin{center}
        \bf Ieva \v{C}epait\.{e}\\                           
        December 2017                                    
\end{center}
\vspace*{5mm}
%
%
\begin{abstract}
        Due to the limited availability of quantum computing power in the near future, cryptographic security techniques must be developed for secure remote use of current and future quantum computing hardware. Prominent among these is Universal Blind Quantum Computation (UBQC) and its variations such as Quantum Fully Homomorphic Encryption (QFHE), which herald interactive and remote secure quantum computing power becoming available to parties that require little more than the ability to prepare and measure single qubits. Here I present a simulation of such a protocol, tested classically on the simulation platform LIQ$Ui\ket{}$ and then later adapted to and run on the recently released IBM 16-qubit quantum chip using their beta cloud service. It demonstrates the functionality of the protocol and explores the effects of noise on potential physical systems that would be used to implement it.
        
\end{abstract}

\subsubsection*{Declaration}

\begin{quotation}
        I declare that this project and report is my own work.
\end{quotation}

\vspace*{0.5cm}
Signature:\hspace*{8cm}Date: 30th November 2017 

\vfill
{\bf Supervisor:} Dr. E. Kashefi               
\hfill                                        
\newpage
%

\pagestyle{plain}                               

\setcounter{page}{1}                            

\tableofcontents                                

\newpage

\section{Introduction}

Quantum computers and quantum algorithms promise to usher in a new era of information processing, solving problems that prove intractable for classical computers. Whether it is the simulation of physical quantum systems\cite{Lloyd1073} or factoring large prime numbers in timescales previously thought to be impossible\cite{shor}, it is clear that quantum processors can and will be applied universally once the technologies for it are fully developed.

The fragile nature of quantum systems- particularly those used in information processing- means that access to these technologies, such as those developed by IBM\cite{ibm16qubits} is likely to be limited. This has led to an interest in $distributed$ quantum computation. In order for individuals and institutions to be able to use these processors remotely, through a cloud-based service for example, there exists a need for security protocols that would allow the computations to be carried out privately, without servers gaining any access to the nature of the information being processed, up to a certain minimal, leaked amount (such as the size of the input or output).

A novel protocol\cite{UBQCmain} for performing remote and secure quantum computations was introduced by Broadbent, Fitzsimons and Kashefi in 2008 called Universal Blind Quantum Computation (UBQC). It has since been shown to be secure and correct, both in a stand-alone setting and as a cryptographic primitive\cite{delegatedcomputations}. Variations on this protocol have been proposed since, such as the Quantum Fully Homomorphic Encryption (QFHE) scheme. UBQC and QFHE will be described in more depth in the next chapter of this report as well as the model of computation they are constructed from- Measurement Based Quantum Computing (MBQC).

The goal of this project was to understand and classically simulate a recently proposed QFHE scheme, developed by P. Wallden, M. Hoban, A. Gheorghiu and E. Kashefi. The simulation was first implemented using Microsoft's LIQ$Ui\ket{}$\cite{liquid} simulation platform and then adapted to be run on the recently released IBM 16-qubit cloud quantum processor in order to verify its outputs and investigate the impact of noise on such protocols where current quantum computing technologies are concerned.

This report will aim to guide the reader through the necessary background on Quantum Computation, MBQC and the QFHE protocol before presenting the methods used for the simulations and the results that they produced.

\section{Background}

\subsection{Quantum Computing}

The analogy to a classical bit in quantum computing is the $qubit$. It is a two-state quantum-mechanical system whose state can be represented as a unit vector in two-dimensional complex Hilbert space. The orthonormal basis in that space is usually labelled with basis vectors $\ket{0}$ and $\ket{1}$. Any state of a qubit can then be represented as a linear superposition of the two bases:
\begin{align}
\alpha\ket{0} + \beta\ket{1},
\end{align}
where $\alpha$ and $\beta$ are complex coefficients with the property $|\alpha|\textsuperscript{2} + |\beta|\textsuperscript{2} = 1$. A tool to geometrically represent the states of a qubit is the Bloch sphere as seen in Figure \ref{bloch}. For a two-dimensional Hilbert space, complex projective lines in the sphere are subspaces that all together represent all pure quantum states of the computer.

\begin{center}
    \includegraphics[width=0.25\linewidth]{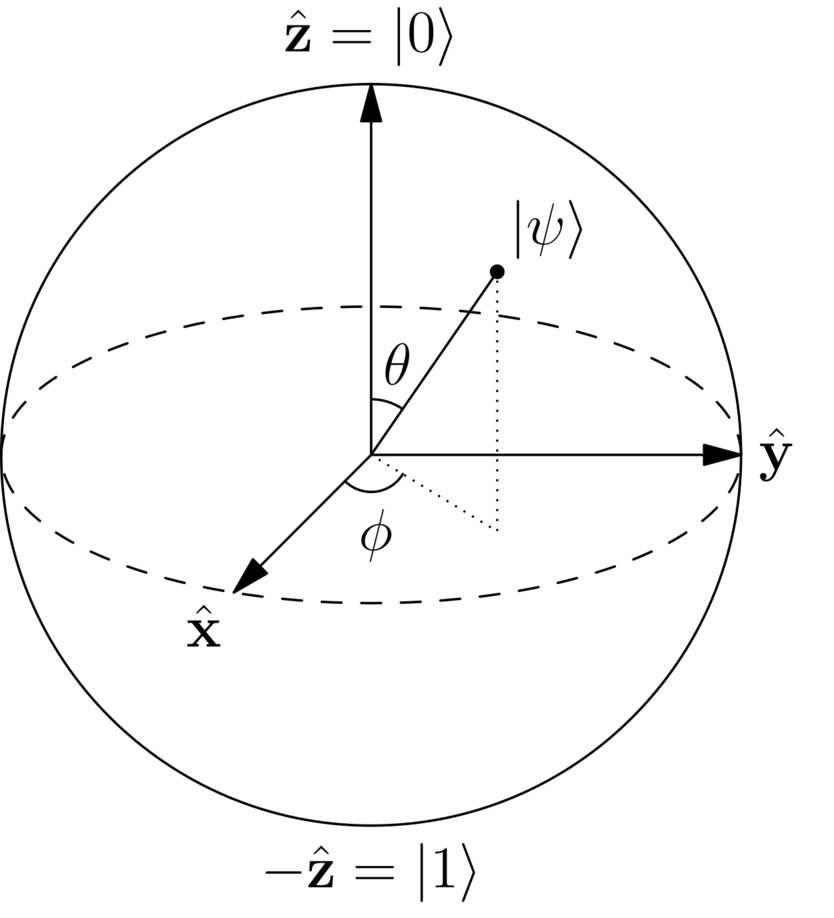}
    \captionsetup{width=0.8\textwidth}
    \captionof{figure}{The Bloch Sphere: a geometrical representation of a two-level quantum mechanical system. Lines on the sphere represent two-dimensional quantum states with the North and South poles usually correspond to the $\ket{0}$ and $\ket{1}$ basis states respectively.}
    \label{bloch}
\end{center}

The basis $\{\ket{0},\ket{1}\}$ is known as the computational basis of the quantum computer and computations are performed by applying unitary operations to it and changing the qubit's state. This is easily visualized as rotating the line of a quantum state on the Bloch sphere at different angles to new points.

The $n$ qubit state is a unit vector in the tensor product space $\mathcal{H}_1 \otimes \mathcal{H}_2 \otimes ... \otimes \mathcal{H}_n$. The $2^n$ basis states of this space are $n$-fold tensor products of $\ket{0}$ and $\ket{1}$, where $\ket{0}\otimes ... \otimes\ket{0}$ is written as $\ket{0...0}$. Thus, any state of the quantum computer $\ket{\psi}$ is a $2^n$ dimensional complex unit vector that can be expressed as:
\begin{align}
\ket{\psi} = \sum\limits_{i\in \{0,1\}^n}\alpha_i\ket{i}.
\end{align}

A unitary operation on this quantum state is usually represented as a quantum gate in a circuit, as seen in Figure \ref{circuit2}. The gates are applied from left to right in  time-steps until the computation is complete. These unitaries are often denoted by matrices that are applied to the state vectors. The gates used in this project and their notation are found in Appendix B.

 Every computation realizable by a quantum circuit can be implemented by using a small number of gates called a $universal$ set. This means that for any integer $n \geq 1$, any $n$-qubit unitary operator can be expressed by a quantum circuit using only a finite number of gates from the universal set.

\begin{center}
    \includegraphics[width=0.7\linewidth]{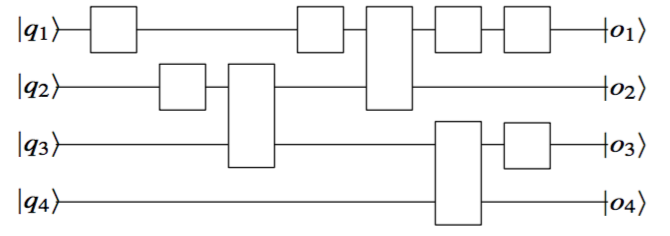}
    \captionsetup{width=0.8\textwidth}
    \captionof{figure}{A generic quantum circuit, where the wires represent single qubits and $\ket{q_i}$ are initial qubit states while $\ket{o_i}$ are the final output states. The rectangles are quantum gates, which can span any number of wires and apply specific unitary operations to the qubits they are drawn on.}
    \label{circuit2}
\end{center}

\subsubsection{Measurement}

While the state of a classical computer can be read at any point during a computation without interfering with the process, observing the state of a quantum computer will immediately collapse its superposition of basis states into a single eigenstate. The measurement of a quantum system is essentially the application of measurement operators $M_i$ with \textit{i} different outcomes on the state of the computer $\ket{\psi}$. The probability of the measurement outcome depends on the pre-measurement state of the system:
\begin{align}
p(i) = \bra{\psi}M_{i}^{\dagger}M_i\ket{\psi},
\end{align}

and the operators $\{M_1,...M_i\}$ have to obey completeness. A special case of measurement operators are projection operators, where $M_i^{\dagger}M_i\ket = P_i$ and they are pairwise orthogonal: $P_iP_j = \delta_{ij}P_i$. After a measurement occurs the state of the system becomes $\frac{1}{\sqrt{p(i)}}P_i\ket{\psi}$ and the change of the system from the state $\ket{\psi}$ to $P_i\ket{\psi}$ is not a unitary evolution but rather the collapse of the wavefunction. Thus measurement operations cannot be undone and lead to difficulties in devising quantum algorithms.

\subsubsection{Noise and Decoherence}

Beyond the issues concerning measurement, a final important thing to note is the concept of \textit{decoherence}. This it the loss of quantum information over time due to the qubit's interaction with the environment. Quantum mechanical systems are incredibly volatile and most qubit technology relies on near-perfect isolation from any disturbances that might change the state of the system. However perfect isolation is impossible and a qubit's state will likely change or 'decohere' due to the time it is left in contact with its environment or the various operations that are performed on it during the computation.\cite{2005quant.ph..5070D} There are quantum codes developed to deal with the problems caused by this\cite{nielsen2000quantum} such as the infamous Shor's Code but, unfortunately, they require even more qubits to be introduced to the circuit, of which there is still a shortage in the world of technology.

\subsubsection{Entanglement and Bell States}

One of the most important quantum mechanical phenomena used in quantum computation is entanglement, best illustrated by the so-called Bell States. These are pairs of qubits that exhibit the Einstein-Podolsky-Rosen Paradox and are thus known as EPR pairs.\cite{perry2012quantum}\cite{nielsen2000quantum} The concept can extend to more than two qubits and they are used extensively in the protocols discussed later in the report. The mechanism of creating a Bell pair of two qubits is shown in the diagram below:
\begin{align*}
\Qcircuit @C=1em @R=.7em {
& \lstick{\ket{x}} \qw & \gate{H} & \ctrl{2} & \qw \\
& \quad & \quad & \quad & \rstick{\beta\textsubscript{xy}}\\
& \lstick{\ket{y}} \qw & \qw & \targ & \qw
}
\end{align*}

\[
    \ket{00} \xrightarrow[]{\beta} \frac{1}{\sqrt{2}}(\ket{00} + \ket{11}) = \beta\textsubscript{00}
\]
\[
    \ket{01} \xrightarrow[]{\beta} \frac{1}{\sqrt{2}}(\ket{01} + \ket{10}) = \beta\textsubscript{01}
\]
\[
    \ket{10} \xrightarrow[]{\beta} \frac{1}{\sqrt{2}}(\ket{00} - \ket{11}) = \beta\textsubscript{10}
\]
\[
    \ket{11} \xrightarrow[]{\beta} \frac{1}{\sqrt{2}}(\ket{01} - \ket{10}) = \beta\textsubscript{11}
\]

The advantage of Bell States is not immediately apparent. They are states of composite systems that cannot be written as tensor product of the states of its components. Computationally, it can be seen that measuring the state of one qubit will immediately tell you the value of the other. In the state $\beta_{00}$ for example, if the outcome of the measurement of the first qubit is 0, then the value of the second qubit can only be 0 too. This will be exploited in the models discussed in the following sections.

\subsection{Measurement Based Quantum Computation}

The content of this section is largely based on lectures by P. Wallden as found in \cite{IQClectures} and the paper by V. Danos, E. Kashefi and P. Panangaden  \cite{unitaries}.

The MBQC model of quantum computation is realised by creating a large entangled state of qubits and then measuring them in different bases as described by angles on the Bloch sphere, in a particular order. This model of computation is equivalent tos the actions of quantum gates. The qubits are thus evolved unitarily one-by-one, where each measurement transmits information to the next qubit through the entanglement and then corrections are applied in the form of changing the measurement angles based on the \textit{flow} of the computation. While the act of measuring a quantum system implies the loss of information, MBQC can still perform universal quantum computation. Furthermore, it can provide certain advantages in terms of architecture, fault-tolerance and cryptographic applications. 

\subsubsection{The Measurement Pattern}

In order to use MBQC we first have to define the $\pm_{\phi}$ basis:
\begin{align}
    \ket{+_{\phi}} &= \frac{1}{\sqrt{2}}(\ket{0} + e^{i\phi}\ket{1}) \\
    \ket{-_{\phi}} &= \frac{1}{\sqrt{2}}(\ket{0} - e^{i\phi}\ket{1}).
\end{align}
We also define the measurement operation $M_i^{\phi}$ which projects the qubit \emph{i} onto one of the states $\ket{\pm_{\phi}}$ and the $\wedge Z$ gate, which is a two-qubit gate used to entangle them. It is also known as the 'controlled-\emph{Z} gate, since it applies a \textit{Z} operation on a target qubit based on the (unmeasured) state of a 'control' qubit as defined in a circuit. Since its operation is symmetrical, it can be applied using either of the two qubits as the control for an equivalent result.

The classical outcome of the measurement on qubit \textit{i} is denoted as $s_i \in \mathbb{Z}_2 $ and we define $s_i = 0$ if the measurement outcome after applying $M_i^{\phi}$ was $\ket{+_{\phi}}$ and $s_i = 1$ for outcome $\ket{-_{\phi}}$. 
In an MBQC computation, all qubits start out in a $\ket{+}$ state and are then entangled into a large cluster state using $\wedge Z$ operations, before being measured by applying the $M_i^{\phi}$ operator with different values of $\phi$ based on the desired computation. These measurements then induce Pauli gate corrections in unmeasured qubits, where a correction can be an \textit{X} or \textit{Z} gate.

To see how these measurements perform the correct computation, we need to introduce the gate set $J(\alpha)$ (the set of generators $\{\wedge Z, J(\alpha); \alpha\in[0,2\pi)\}$ is universal and can approximate any transformation in $\mathbb{C}^2$, as discussed in \cite{unitaries}). A $J(\alpha)$ gate can then be turned into a one-way measurement pattern for any $\phi$. If we define $\wedge Z_{ij}$ as the $\wedge Z$ gate being applied between qubits \textit{i} and \textit{j} and $X^{s_i}_j$ as a correction on qubit \textit{j} based on the measurement outcome $s_i$, where $s_i = 0$ implies no correction while $s_i = 1$ would lead to applying an \textit{X} gate to the indicated qubit to rotate it back to the required state. The pattern can now be written as:
\begin{align}
J(\alpha) = X^{s1}_2M_1^{-\alpha}\wedge Z_{12}.
\end{align}
The above consists of entangling two qubits, measuring the first one at angle $-\alpha$ and then applying a correction on the second qubit based on the outcome of the first. The second qubit is then the 'output' that carries the information of the applied gate onto the next layer of measurements.

\subsubsection{Flow and Graphs}

The flow is an important construct that defines the order of measurements and corrections in MBQC. First we need to consider an \textit{open graph state} (G, I, O) which consists of an undirected graph \textit{G} as well as subsets of nodes \textit{I} and \textit{O} which are known as inputs and outputs respectively, with $I^c$ and $O^c$ being defined as their complements. It is said to have \textit{flow} if there exists a map $f : O^c \rightarrow I^c$ and a partial order $\preceq$ over qubits\cite{flow}:

\begin{itemize}
    \item $x \sim f(x):$ \textit{x} and \textit{f(x)} are neighbours on the graph
    \item $x \preceq f(x):$ (\textit{f(x)} is to the future of \textit{x} with respect to the partial order)
    \item For all $y \sim f(x)$, we have $x \preceq y:$ (any other neighbours of \textit{f(x)} are to the future of \textit{x})
\end{itemize}

\begin{center}
    \includegraphics[width=0.4\linewidth]{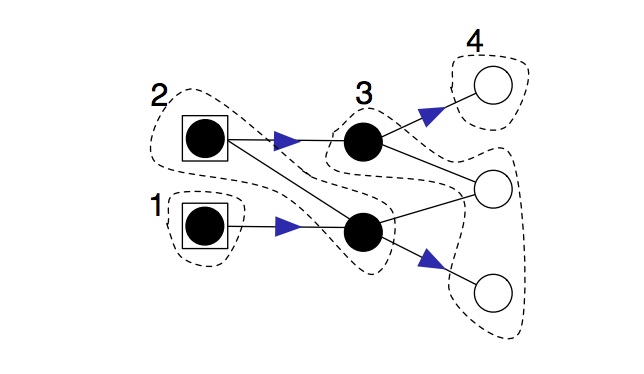}
    \captionsetup{width=0.8\textwidth}
    \captionof{figure}{An open graph state with flow. The boxed qubits are the inputs and white circles are the outputs. All the non-output qubits, black circles, will be measured during the run of the pattern. The flow function is represented as arrows and the partial order on the vertices are given by the 4 partition sets.\cite{flow}}
    \label{flow}
\end{center}

The \textit{X} and \textit{Z} Pauli gates have properties:

\begin{align}
M_i^{\phi_i}X &= M_i^{-\phi_i} \\
M_i^{\phi_i}Z &= M_i^{\phi_i + \pi}
\end{align}

So for qubit \textit{i} the corrections can be applied by changing the measurement angle and accumulate as:

\begin{itemize}
    \item an \textit{X} correction from $f^{-1}(i)$, the qubit whose flow is \textit{i}
    \item a \textit{Z} correction from all qubits $j \neq i$ whose flow $f(j)$ is a neighbour to \textit{i} 
\end{itemize}

The actual measurement angles for qubits in the graph is then defined as:
\begin{align}
\phi_i' = (-1)^{s_{f^{-1}(i)}}\phi_i + \pi(\sum\limits_{j:i\in N_G(f(j)), j\neq i}s_j),
\end{align}
where $N_G(f(j))$ are all the neighbours of $f(j)$ in the graph.

These MBQC Graph states are often represented in diagrams where edges are entanglements and vertices are qubits as in Figure \ref{flow}. To represent flow in MBQC computations, the graph states are often regular in shape and read from left to right, which is particularly important in large \textit{brickwork states} as in Figure \ref{brickwork}, which are constructed when MBQC is performed for UBQC, as discussed in the next section.

\subsection{Quantum Fully Homomorphic Encryption Scheme}

Now that the groundwork is set for the MBQC model, it is possible to introduce how it is applied in potential Quantum Encryption schemes and protocols. The most prominent of these is UBQC, which is discussed briefly below. However, many variations have been constructed since, such as the QFHE protocol that was simulated in this project. 

\subsubsection{Universal Blind Quantum Computation}

UBQC was developed to allow a client (Alice) with no quantum computational power or memory to be able to delegate her computations to a quantum server (Bob) completely privately, without leaking knowledge about the input, output or even the computation itself. Alice's abilities would have to consist of being able to prepare single qubits with pre-rotated angles that Bob would then have no knowledge of.

\begin{center}
    \includegraphics[width=0.7\linewidth]{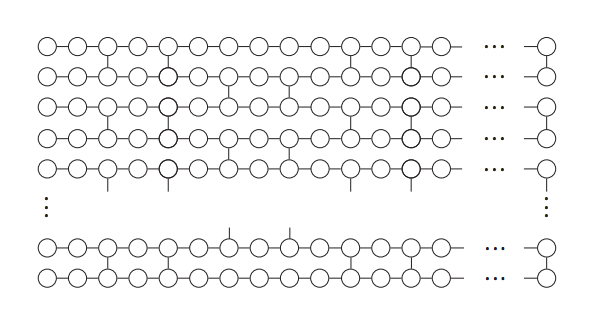}
    \captionsetup{width=0.8\textwidth}
    \captionof{figure}{A graph of the \textit{brickwork state} as constructed for UBQC computations. \cite{UBQCmain}}
    \label{brickwork}
\end{center}

Bob would then have to entangle these qubits in a pre-determined brickwork state, as shown in Figure \ref{brickwork} and discussed in \cite{UBQCmain}. This would inevitably reveal the upper bounds of the dimensions of the graph to Bob, but no information is available to them beyond that. The brickwork state can be created for any universal quantum gate by padding it with extra qubits so as to hide the shape of the graph.

Finally, Bob would begin measuring the qubits in layers according to the given flow and send the results of their measurements to Alice, who would then communicate the angles at which to measure the next layer of qubits. These would be computed based on the position of the qubit in the brickwork state and the padding she adds to each measurement by pre-rotating the qubits she sends at angles chosen from some set, as well as adding a single one-time $\pi$ rotations in a randomized fashion. Thus when she gets the results, she can apply corrections to the outputs to decode the results of her computation. 

She can also verify that the computation being performed is correct by hiding 'dummy' qubits in the brickwork whose output she would know beforehand. Thus the computation would be private and correct provided the noise has no impact. While this is a very brief discussion of UBQC, a detailed description can be found in \cite{UBQCmain}.

\subsubsection{The Simulated QFHE Scheme}

While UBQC is a powerful scheme, it also requires a high level of interaction between the classical and quantum machines as well as a very large number of qubits to create the brickwork state. Some of these issues are addressed with a different type of encryption scheme: the QFHE. Although the protocol is as of yet unpublished, this section will briefly summarise the theory behind it as developed by P. Wallden, M. Hoban, A. Gheroghiu and E. Kashefi.

The idea behind QFHE is that Bob can perform an arbitrary quantum computation on data while being aware of the computation but unaware of the inputs or outputs themselves. As in UBQC, Alice prepares her single qubits and instructs Bob to entangle them. However in this case all of them are in the $\ket{+}$ state and the entanglement does not need to be a brickwork state as she does not care about revealing the nature of her computation and can save on the extra resource qubits. She also sends an extra qubit for every qubit measured at angle $\phi_i \in \{\frac{\pi}{4}\}$ and instructs Bob to create a Bell Pair using them.

Bob then performs the measurements as per the MBQC pattern but does not measure each one qubit of the Bell Pairs and instead sends them back to Alice so she could measure them herself. This requires Alice to have the quantum resources to measure single qubits, but nothing beyond that.

This is because the computation lacks the interactive measurement of UBQC and instead Alice performs all of her corrections at the end of the process using the fact that Pauli corrections \textit{X} and \textit{Z} commute with all Clifford operations (operations that can be simulated efficiently on a classical computer) so after each measurement the correction is teleported through to the next qubit and so all the way to the very end. This includes her classical input, meaning she can hide whether the input qubit is $\ket{0}$ or $\ket{1}$ by performing a correction at the end:
\begin{align}
101 \rightarrow \ket{-}\ket{+}\ket{-} \rightarrow Z^1_1Z^0_2Z^1_3\ket{+}\ket{+}\ket{+}.
\end{align}
This does not, however, hold true for \textit{T}-gate operations, which require a measurement at angles of $\frac{\pi}{4}$ (non-Clifford):
\begin{align}
TX^aZ^b = X^aZ^{a \oplus b}S^aT.
\end{align}

An unwanted S-gate appears after the corrections are teleported and it needs to be removed. This is done by measuring Alice's Bell Pair qubit in either the X or Y-basis depending on the outcome of Bob's Bell Pair qubit and then using that as a correction as discussed in the next section.

Thus Bob remains unaware of the inputs and does not know all the corrections for the outputs.

\subsubsection{Corrections}
This section gives a quick overview of the corrections on each qubit based on the measurements before it due to the flow of the measurement pattern. $s_i$ are measurement outcomes, $b_i$ are corrected outcomes, calculated in increasing order. $\alpha_i$ is the measurement outcome of Alice's qubit in each Bell pair in the case that there is a qubit in the computation measured at  $\frac{\pi}{4}$. The flow is represented by $f(\cdot)$ and $Z^i$ is the set of qubits that have a Z-dependence on qubit $i$. This means that qubits $j\in Z^i$ have to have a Z-correction based on the value of $i$. These corrections extend to the input, meaning there should be an extra layer of corrections depending on whether or not the value of the input is 1.

Bob always has to measure in default, non-corrected angles, while Alice has to measure her extra qubit in a different basis:
\begin{align}
\beta_i =b_{f^{-1}(i)},
\end{align}
where $\beta_i = 0$ means Alice has to measure in the X-basis and $\beta_i = 1$ means measuring in the Y-basis. 

If the measurement angle in the MBQC pattern is taken to be $\phi_i$, then the following corrections should be performed:

If $\phi_i \in \{0, \pi\}$ then:
\begin{align}
b_i = s_i \oplus \sum\limits_{j\in Z^i}b_j.
\end{align}

If $\phi_i \in \{\frac{\pi}{2}, \frac{3\pi}{2}\}$ then:
\begin{align}
b_i = s_i \oplus b_{f^{-1}(i)} \oplus \sum\limits_{j\in Z^i}b_j.
\end{align}

If $\phi_i \in \{\frac{\pi}{4}\}$ then:
\begin{align}
b_i = s_i \oplus \alpha_i \oplus \sum\limits_{j\in Z^i}b_j.
\end{align}

\section{Method}

The aim of this project was to simulate a simple application of the QFHE protocol both classically and then on the IBM 16-qubit chip once beta-access was granted for its use. The classical simulations were first coded in Microsoft's LIQ$Ui\ket{}$ simulation platform for an efficient test of correctness, then adapted to the IBM Python API and its chip's hardware. 

The most important aspect of the simulation was to show that a given quantum computation would produce the same probability amplitudes for each qubit with interactive corrections as with the deferred corrections and Bell pair measurements done by Alice. As discussed above, the Pauli corrections are reliant on the flow of the qubit measurement patterns\cite{flow}, thus the circuits that were simulated had to have $\frac{\pi}{4}$ measurements at particular locations on the graph state in order to have an impact on the final corrected qubit amplitudes.

This means that most of the MBQC circuits derived from circuits originally composed of unitaries such as Pauli gates and \textit{T}-gates would often not be as optimal as simply implementing a specially designed MBQC computation, where the position of qubits in the graph state and the angles at which they were measured could more easily be tailored to fit the parameters to be explored. The final circuit does not perform any 'useful' computation but rather works as a means to graphically and numerically verify that the protocol performs as it should.

An important consideration was the size of the circuit. Classical simulations of quantum systems require large amounts of memory and processing power in order to deal with the number of different states that are being operated on at once. This means that the circuit cannot be composed of too many qubits, else the classical simulation would not run without the help of a supercomputer. A classical computer can usually handle up to about 15-30 qubits, depending on the type of and efficiency of the implementation as well as the computer's processing power. Beyond that, however, the maximum number of qubits that could be used in this simulation would always be 16 in order to later implement it on the IBM chip. In fact, the number had to be even lower because of the adjustments and possible extra qubits required for the final implementation on the IBM machine as discussed later in the section.

After a series of trials and errors, the graph of the final circuit which was simulated can be seen in Figure \ref{circuit}.

\begin{center}
    \includegraphics[width=0.3\linewidth]{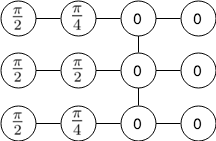}
    \captionsetup{width=0.5\textwidth}
    \captionof{figure}{A graph of the final MBQC circuit used in the simulation. The angles on each vertex represent the angles of measurement of each qubit and the flow is from left to right.}
    \label{circuit}
\end{center}

\subsection{LIQ$Ui\ket{}$}

LIQ$Ui\ket{}$ is a simulation platform for quantum computation that operates in the F\# language. Its high functionality allowed for easy and efficient implementation of different families of unitaries and models of quantum computation and would run simulations much faster than the IBM Python API. It was used to test many different candidate circuits. 

Most of the tests consisted of starting out with a simple quantum circuit, then testing whether a conversion to the gate set $J(\alpha)$ would affect the amplitude outpts of the simulation. The reason this was done is because the conversion from a computation composed of these unitaries to an MBQC pattern would serve as an intermediary test for correctness when switching between quantum circuits and MBQC patterns.

For each different type of simulation, the circuit was run 1000 times and corrected output values for each of the 3 qubits in the output layer were collected as statistics and plotted as histograms using in-built F\# methods to visualise the final probability amplitudes for each qubit after the computation. The input qubits could classically represent values 0-7 and the results were plotted for all of those inputs.

A full manual for the simulator can be found at \cite{liquidmanual} and examples of code are listed in Appendix A.

\subsubsection{Interactive Corrections}

The first version of the circuit to be tested had its Pauli corrections written out explicitly as modifications of the measurement angles based on qubits measured in the past as per the flow of the computation (see Section \textbf{2.2.2}). Since LIQ$Ui\ket{}$ allows access to the values of single-qubit measurement outcomes as soon as the measurement is performed, it was possible to simulate the 'interactive' angle modifications to correct outcomes of all the qubits. 

Thus all that was required was to prepare all the qubits as $\ket{+}$ states by applying Hadamard gates to qubits in $\ket{0}$ states, then entangling them with $\wedge Z$ operations as in Figure \ref{circuit} and then measuring them at angles conditioned on previous measurement outcomes (by using the definition of the modified angle from MBQC) or different classical inputs. The code for the measurement part of the circuit can be found in Appendix A.

\subsubsection{Deferred Corrections and Implementing the Protocol}

The next step was to remove the interactive correction method and to attempt to perform the corrections after all the qubits had been measured. The code would then only perform  $M_i^{\phi}$ for the $\phi$ that were pre-determined to perform the computation without changing them based on outcomes obtained during the run. In addition, there would be two extra client qubits added for each of the $\frac{\pi}{4}$ measurements and creating Bell Pairs between them as described in Section \textbf{2.1.1}.

The client halves of the Bell Pairs would then be measured after all the server measurements were done. Finally, results for the output layer of qubits would be calculated by summing-modulo-2 all the relevant measurement outcomes for each output qubit as per the protocol. Examples of this part of the code can be found in Appendix A.

In order to check that without the corrections the server would glean no information about the computation results, values of output qubits were plotted without using any corrections.

\subsection{IBM Quantum Information Software Kit (QISKit)}

This year IBM made strides to release a brand new API using the Python language to allow its clients to compose and run quantum circuits both on simulators and their 5-qubit quantum chip hardware\cite{qiskitapi}. In September, the 16-qubit chip was made available for beta-testing using their cloud server. It was now possible to simulate the QFHE scheme on a real quantum computer.

There were several issues with doing this, as the chip does not allow for networked computations, re-usability of qubits or several other functions necessary or helpful to both carry out the simulation and do it with sufficient fault-tolerance. The methods to adapt the circuit used in LIQ$Ui\ket{}$ and improve its fault-tolerance through various optimization methods are described below.\footnote{More information on the QISKit software is available in their github repository: \url{https://github.com/QISKit}[Last accessed 24/11/2017]}

\subsubsection{Conditional Measurements}

While the LIQ$Ui\ket{}$ simulator allows for a lot of freedom in extracting measurement values throughout the run of a circuit as well as an easy simulation of interactive computation, that is not the case of the IBM server. This is due to the fact that a computation has to be done on a single chip whose functionality only allows for a final readout of the state of the system, rather than single values of qubits throughout. This creates a problem when simulating the protocol, as Alice's measurements of her Bell pair qubits are conditional on the results of the measurements that Bob has supposedly already performed and transmitted to her.

Luckily, there is a relatively simple workaround which complicates the computation only by adding a few unitaries and does not require any additional qubits. This modification, in its simplest form, is a controlled phase gate $\wedge S^{\dagger}$. As discussed in section \textbf{2.4.1}, we know that Alice's qubit has to be measured in a basis 
\begin{align}
\beta_i =b_{f^{-1}(i)},
\end{align} 
with $\beta_i = 0$ being a measurement in the X-basis and $\beta_i = 1$ a measurement in the Y-basis. In essence her basis depends on all the corrections applied to the qubit $b_{f^{-1}(i)}$. In the circuit used for the simulation (Figure \ref{circuit}), each $\beta_i$ is determined from a qubit measurement outcome that does not require corrections, since they are in the first layer of qubits in the flow- the input layer. The only possible X-corrections they might require are from the classical input value of the qubit and these can be applied as X-gates once the rotation of the qubits to the required basis was performed. \footnote{Note that if there were multiple corrections from $i$ different qubits, the simple solution would be to add up their contributions by using $CNOT$ gates with the correction-requiring qubit as the target. If its initial measurement outcome were $b_i$, each of its corrections would come from the qubits with measurement outcomes $s_i$. Thus the corrected measurement would be  $\beta_i = b_i \oplus s_1 \oplus s_2 \oplus ... \oplus s_i$.}

A measurement in the Y-basis is the same as performing a rotated measurement with angle $\phi = -\frac{\pi}{2}$. Thus we can use the qubit determining $\beta_i$ as a control qubit that performs a $-\frac{\pi}{2}$ rotation on Alice's qubit if its value is 1 and does nothing if it is 0. This is a gate that can be constructed using the relation\cite{verification}:
\begin{align}
T^{\dagger}XTX = e^{i\pi/4}S.
\end{align}
\newline

Normally a global phase induced on a state (such as the factor of $e^{i\pi/4}$ in front of the $S$ gate) has little impact in the computation, since the observable value would remain unchanged, but since this is a controlled operation, an outcome 1 of the control qubit would give $e^{i\pi/4}S$ on the target instead of the $S$ operation that we needed. Due to this, we need to add an additional $T^{\dagger}$ gate to the control qubit, or in our case a $T$ gate since the final operation we are hoping to get is $S^{\dagger}$:
\begin{align*}
\Qcircuit @C=1em @R=.7em {
&\ctrl{2} & \qw & \quad & \quad & \quad & \gate{T} & \ctrl{2} & \qw & \ctrl{2} & \qw \\
& \quad & \quad & \quad & \lstick{=} \quad & \quad & \quad & \quad & \quad & \quad  \\
&\gate{S^\dag} & \qw & \quad & \quad & \quad & \gate{T^\dag} & \targ & \gate{T} & \targ & \qw 
}
\end{align*}

So Alice's qubit measurement is now done at the correct basis. The only thing left to take care of is to apply the corrections from the relevant qubits to the output layer, which is done using $CNOT$ gates. These are essentially controlled-\textit{X} gates with the control being the qubit inducing a correction  and the target being the qubit whose value needs to be flipped with the \textit{X} operation. Hence the only measurements performed in this modified model are on the qubits in the output layer, with the others being rotated to the correct measurement basis without actually performing any measurements operations. 

\subsubsection{Qubit Couplings}

Once the circuit is rewritten to be run on a single chip, the issue of connectivity also needs to be resolved. The IBM 16-qubit transmon chip has a specific topology that cannot be re-arranged (see Figure \ref{ibmchip}), so any entanglements can only happen between qubits that are in the correct positions on the connectivity graph.

\begin{center}
    \includegraphics[width=0.8\linewidth]{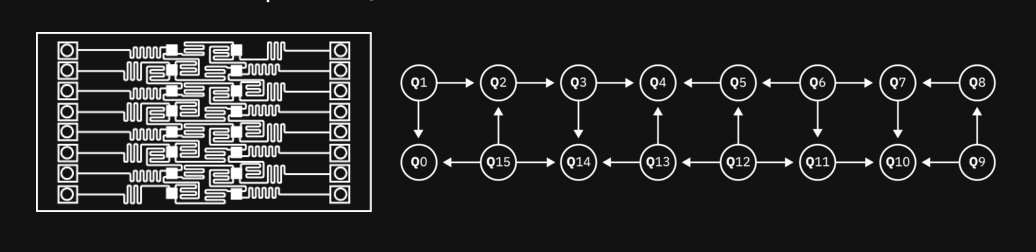}
    \captionof{figure}{A schematic of the IBM 16-qubit chip (left) and a coupling map describing all the possible two-qubit gate connections (right), where the arrow points from the control-qubit to the target.}
    \label{ibmchip}
\end{center}

The solution was to optimize the arrangement of all the qubits to bring the pairs requiring two-qubit gates as close together as possible and then implement SWAP-gates where required to ensure the direction of the operation and the required qubits were in the right place. The usual implementation of a SWAP gate would require a two-way control channel, but there is a modification that implements a SWAP using only one control-qubit:
\begin{align*}
\Qcircuit @C=1em @R=.7em {
&\ctrl{2} & \targ & \ctrl{2} & \qw & \quad & \quad & \quad & \ctrl{2} & \gate{H} & \ctrl{2} & \gate{H} & \ctrl{2} & \qw \\
& \quad & \quad & \quad & \quad & \quad & \lstick{=} & \quad & \quad & \quad & \quad & \quad & \quad & \quad \\
&\targ & \ctrl{-2} & \targ & \qw & \quad & \quad & \quad & \targ & \qw & \gate{Z} & \qw & \targ & \qw
}
\end{align*}

The connectivity of the chip and these methods were enough to arrange the circuit in a way that would allow for its successful execution.

\subsubsection{Final Optimization and Classical Post-Quantum Processing}

The above methods, while replicating the operations of the QFHE protocol, also increase the complexity of the circuit and the possibility of decoherence of the qubits as well as errors induced on them by applying gates. Thus a final set of runs for the simulation was done with a number of optimisation methods and using classical post-processing of measurements as in the LIQ$Ui\ket{}$ case, since the Python language API that IBM has provided would allow for some access to single qubit measurements.

The optimisation of gates consisted of removing as many unnecessary SWAP operations as could be afforded (such as in the creation of Bell Pairs, where the two qubits in the pair would be identical for all intents and purposes) and removing unnecessary H operations and converting $\wedge Z$ operations into CNOTs. This was due to the fact that a $\wedge Z$ operation could effectively be decomposed into H and CNOT gates, which is precisely how the IBM hardware would operate since its hardwired gate-set did not include $\wedge Z$ unitaries. This often led to two H gates being applied to a qubit in succession, which was effectively just applying an Identity gate. Thus they could both be removed as long as the rest of the circuit was modified to not be affected, such as in the example below:
\begin{align*}
\Qcircuit @C=1em @R=.7em {
& \qw &\ctrl{2} & \qw & \quad & \quad & \quad & \qw & \qw & \ctrl{2} & \qw & \qw & \quad & \quad & \quad & \ctrl{2} & \qw & \qw \\
& \quad & \quad & \quad & \quad & \lstick{=} \quad & \quad & \quad & \quad & \quad & \quad & \quad & \quad & \lstick{=} & \quad & \quad & \quad & \quad \\
& \gate{H} & \gate{Z} & \qw & \quad & \quad & \quad & \gate{H} & \gate{H} & \targ & \gate{H} & \qw & \quad & \quad & \quad & \targ & \gate{H} & \qw 
}
\end{align*}

Besides the optimisation of the circuit itself, the possibility of extracting single-qubit measurements and ridding the computation of a number of SWAP and CNOT gates which were used to apply error corrections was also implemented. The IBM QISKit API outputs its computation results as a dictionary of final measured states of the chip and the number of times that the particular state was measured out of the number of times that the circuit was run. Through a series of trials and errors, it was possible to determine which bit in the resulting outputs corresponded to a particular measured qubit value and how they were arranged. Using this, a classical method was implemented to correct the measured outcomes as in the LIQ$Ui\ket{}$ code. This required more measurements to be performed by the hardware, but reduced the number of gates and hence the runtime by a significant factor. As in LIQ$Ui\ket{}$, all the error-inducing qubits were measured as well as the layer of output qubits. These were then summed-mod-2 in the right combinations to produce the required final output values of the computation. 

\section{Results and Discussion}

This section presents the numerical and graphical data was obtained from the simulations and runs on the IBM chip using the methods discussed above. It also discusses the meaning of these results in terms of the QFHE protocol as well as the hardware performance of the 16-qubit chip.

\subsection{LIQ$Ui\ket{}$ Outputs}

To verify that the protocol performed as it should, simulations on LIQ$Ui\ket{}$ were run a number of times while collecting different statistics from the points of view of Alice and Bob. This involved an unsecure computation where both were privy to the correct output (interactive corrections), as well as QFHE computations where Alice would apply the corrections to her outputs and Bob would simply get uncorrected values. 

The results of some of these runs are presented below in Figures \ref{simulations} and \ref{bobs} as histograms produced using LIQ$Ui\ket{}$'s graphing tools. Although they are each illustrative of only a single experiment, repetitions showed a standard deviation of less than 1\%, so they can be considered as representative of the entire set of 20 runs that were performed.

\begin{figure}[H]
\begin{subfigure}{.5\textwidth}
  \centering
  \includegraphics[width=0.95\linewidth]{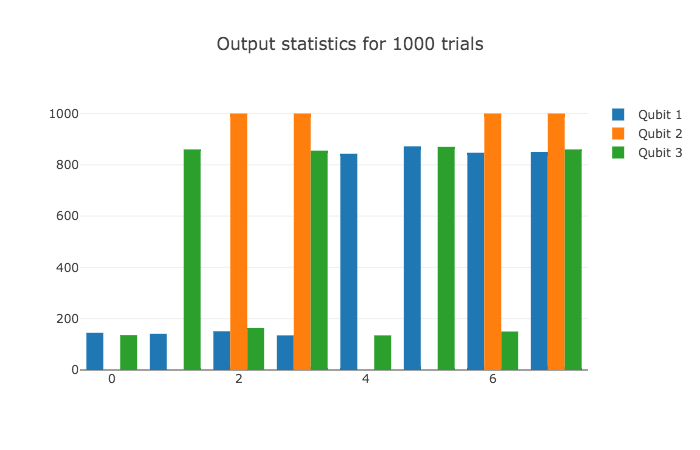}
    \captionof{figure}{}
\end{subfigure}%
\begin{subfigure}{.5\textwidth}
  \centering
  \includegraphics[width=.95\linewidth]{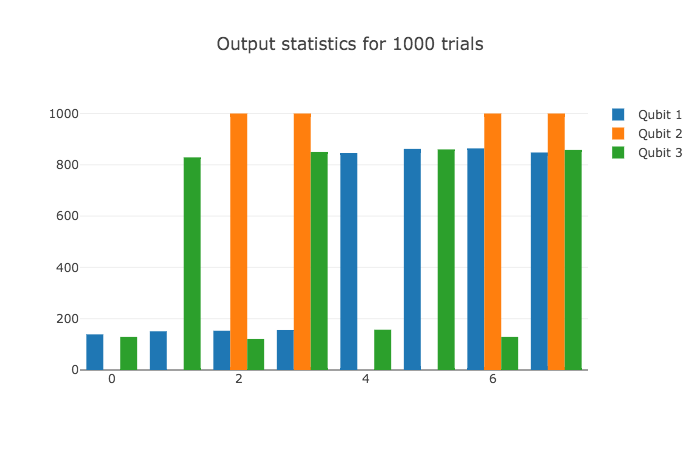}
  \caption{}
\end{subfigure}
\caption{Graphs showing the statistical output of qubit values (y-axis) for each classical input of the computation (x-axis). The graph on the left (a) is produced using the QFHE protocol and applying corrections post-measurement where only Alice would attain the outputs shown, while the graph on the right (b) is one with interactive corrections where both Alice and Bob are aware of the inputs and outputs.}
\label{simulations}
\end{figure}

The histograms in Figure \ref{simulations} are used to show that Alice gets the correct amplitudes of her qubits at the end of the computation, since her corrected values in (a) are significantly similar to the original unsecure computation (b). While the probabilistic nature of the computation outcomes means there is a chance she won't get the same answer on a single run, the simulation is not required to show this, since the original circuit is not meant to be a 'useful' computation. It is enough that the computations are significantly statistically equivalent, so as to demonstrate that if a different computation with a useful result were performed using the QFHE, it would still produce the same outcomes as doing it with an unsecure channel. If the results then need to be verified by running a circuit a number of times as was done above, then this is up to Alice.

\begin{center}
    \includegraphics[width=0.4\linewidth]{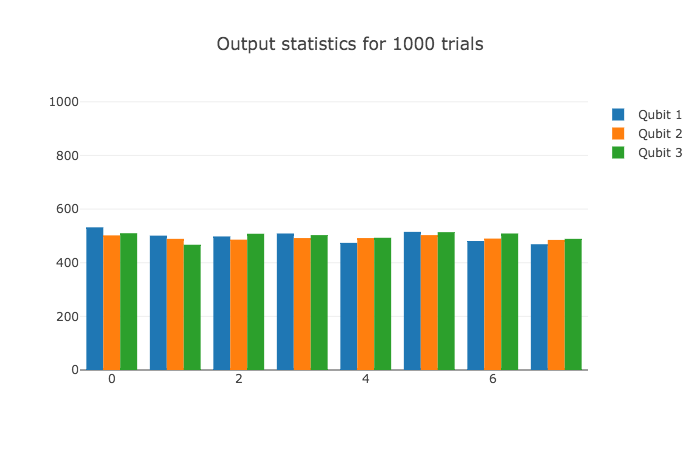}
    \captionof{figure}{Bob's statistical output (y-axis) for each classical input (x-axis) if all he has access to are the performed qubit measurement outcomes and no knowledge of the corrections or Alice's Bell Pair measurement outcomes.}
    \label{bobs}
\end{center}

Figure \ref{bobs} demonstrates that without knowing the outcomes of Alice's Bell Pair qubits or the corrections to be applied due to the input layer, all Bob can really know about the computation are the  outcomes of the uncorrected output qubits, which tell him nothing about what Alice actually obtained as a result.

Without the corrections, the values of output qubits are essentially random, hence they have an almost equal chance of being measured as a 0 or a 1.

\subsection{IBM Simulator Outputs}

The function of running the IBM simulator despite already having simulated results was to check that the translation of what was done with LIQ$Ui\ket{}$ to the IBM platform was successful, since the two work differently in their simulation implementations and the methods provided. IBM's simulator would also predict how the circuit would be run on the real chip and so correctness was paramount.

\begin{center}
    \includegraphics[width=0.5\linewidth]{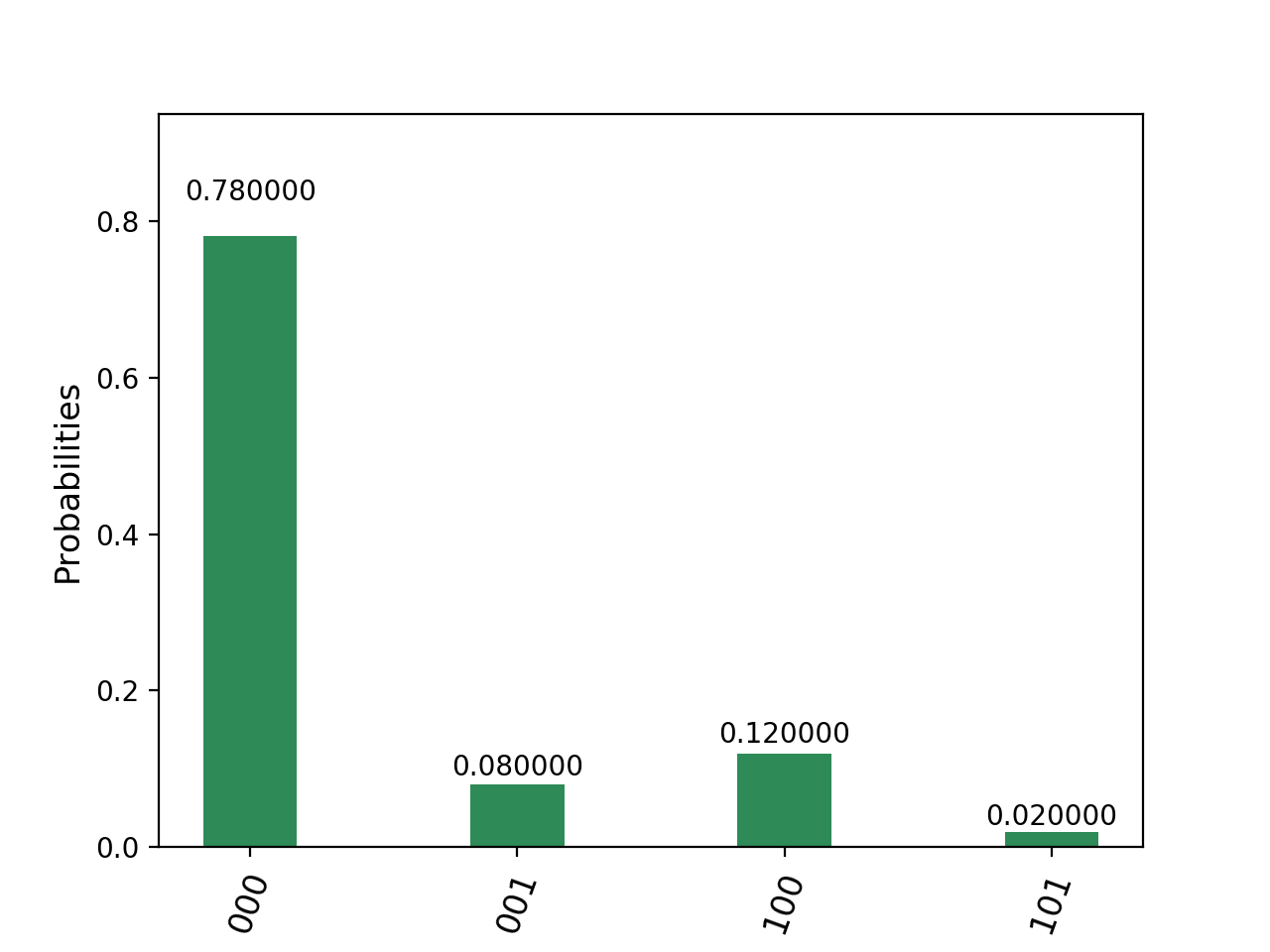}
    \captionsetup{width=0.8\textwidth}
    \captionof{figure}{A histogram showing statistical results for 1000 runs on the IBM simulation of the protocol for classical input 0. It only shows the states that were measured at least once and does not provide information on single-qubit measurements for each separate run, but rather how often a final 3-qubit state as a probability.}
    \label{ibmsim}
\end{center}

The classical inputs would have to be run one-by-one to reflect the performance of the real chip. Figure \ref{ibmsim} is an example of the type of graphical representation of output data that QISKit would provide. It is clear to see, for example, that the middle qubit's value in the final measurements was always 0, since no state was ever measured where its value was 1. But other than that, the API does not provide information on single-qubit values in separate runs of the circuit. 

Single qubit measurements had to be extracted from the simulation runs by processing the numerical data for the final output states and their frequency, which is the greatest level of access to the results of computations that the API allows for. This enabled for a better comparison to the outputs produced by LIQ$Ui\ket{}$ and were thus far more useful than the histograms that QISKit's internal methods produced. Table \ref{simulatortable} shows the counts with post-classical processing of the output data, where an experiment with 1000 runs of the simulator for each classical input value was performed. Again, althogh the experiment was repeated 20 times, the standard deviation of the count values was less than 1\% so the results for this run can be considered as representative for all the repetitions.

\begin{table}[H]
\centering
\begin{tabu}{|p{0.5cm}|[2pt]p{0.5cm}|p{0.5cm}|p{0.5cm}|p{0.5cm}|p{0.5cm}|p{0.5cm}|p{0.5cm}|p{0.5cm}|}
\hline
\multirow{2}{*}{} & \multicolumn{8}{c|}{Classical input value} \\ \cline{2-9} 
                  & 0 &  1 & 2 & 3  & 4 & 5 & 6 & 7\\ \tabucline[2pt]{-}
Q1                  & 128 & 125 & 134 & 140 & 890 & 836 & 854 & 810 \\ \hline
Q2                  & 0 & 0 & 1000  & 1000 & 0 & 0 & 1000 & 1000\\ \hline
Q3                  & 130  & 843 & 140 & 812 & 147 & 855 & 127 & 861 \\ \hline
\end{tabu}
\caption{The number of times out of 1000 runs that a measured value of 1 was observed for each output qubit after the results were processed. These very clearly reflect the corrected outputs obtained in LIQ$Ui\ket{}$.}
\label{simulatortable}
\end{table}

A plot of the above was made using F\#, in order to have a visual representation of the statistical similarities between LIQ$Ui\ket{}$'s results and those of the IBM. It can be seen from Figures \ref{simulations} and \ref{ibmsim} that the computation is performing correctly.

\begin{center}
    \includegraphics[width=0.5\linewidth]{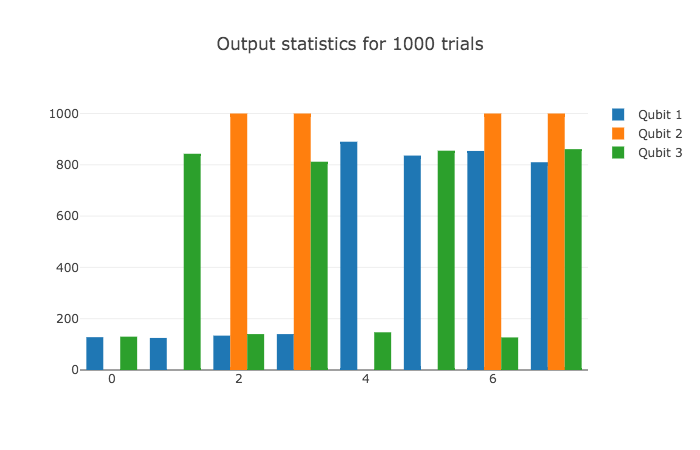}
    \captionsetup{width=0.8\textwidth}
    \captionof{figure}{The results shown in Table \ref{simulatortable} represented by an F\# histogram. The computation results appear to be correct when compared to the output amplitudes obtained in LIQ$Ui\ket{}$.}
    \label{ibmsim}
\end{center}

\subsection{IBM 16-Qubit Chip Outputs}

Once correctness was established using the simulator, a real run could be done on the 16-qubit chip. Each separate classical input was again run 1000 times to collect data and classical post-processing extracted the values for the relevant output qubits which are plotted in Figure \ref{realrun1}.

\begin{center}
    \includegraphics[width=0.5\linewidth]{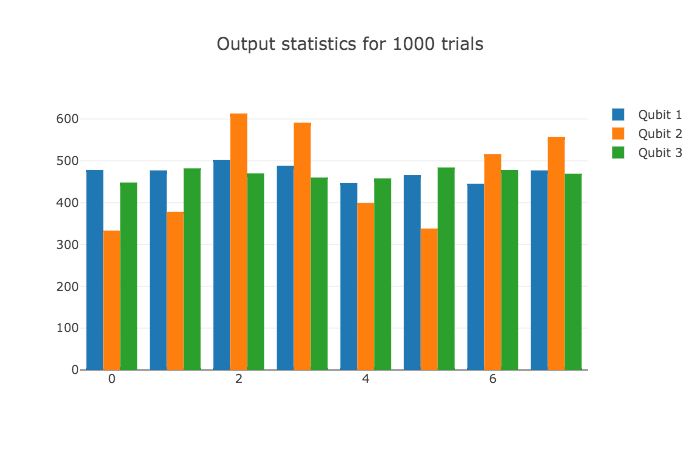}
    \captionsetup{width=0.8\textwidth}
    \captionof{figure}{Results of all initial classical input runs on the 16-qubit chip as plotted with F\#. The effects of decoherence are apparent.}
    \label{realrun1}
\end{center}

It is clear that the effects of decoherence are significant. The statistical pattern that was apparent in the simulation is almost non-existent here, with the qubit amplitude probabilities generally close to $0.5$ on average. This is made more apparent when looking at the numerical values of the output qubits as displayed in Table \ref{initrealruntable}.

\begin{table}[H]
\centering
\begin{tabu}{|p{0.5cm}|[2pt]p{0.5cm}|p{0.5cm}|p{0.5cm}|p{0.5cm}|p{0.5cm}|p{0.5cm}|p{0.5cm}|p{0.5cm}|}
\hline
\multirow{2}{*}{} & \multicolumn{8}{c|}{Classical input value} \\ \cline{2-9} 
                  & 0 &  1 & 2 & 3  & 4 & 5 & 6 & 7\\ \tabucline[2pt]{-}
Q1                  & 478 & 477 & 502 & 488 & 447 & 466 & 445 & 477 \\ \hline
Q2                  & 333 & 378 & 613 & 591 & 399 & 338 & 516 & 557\\ \hline
Q3                  & 448 & 482 & 470 & 460 & 458 & 484 & 478 & 469 \\ \hline
\end{tabu}
\caption{The number of times out of 1000 runs that a value of 1 was measured on the 16-qubit chip. Qubits 1 and 3 appear to have completely decohered, with their amplitudes being around 0.5 on average. The middle qubit still appears to show some variation but it is not nearly significant enough to be treated as a useful result.}
\label{initrealruntable}
\end{table}

The average deviation from the values in the simulation was calculated to be around 37.2\%. This is disappointing and yet not unexpected, as IBM provides data on the average errors of all operations for each qubit on the chip\footnote{\url{https://quantumexperience.ng.bluemix.net/qx/editor}[Last accessed 26/11/2017]}. These seem to imply that the magnitude of the error is to be expected for a circuit with over 100 unitary operations, many of which are applied between 2 qubits. The average error for a single gate is predicted to be of the order of $10^{-3}$ while multi-qubit gates and read-outs will cause errors of magnitude $10^{-2}$. This, together with the time it takes for the computation to be completed (during which the qubits continue to decohere until they are measured), is likely to cause enough of an error to just reduce each qubit's measured value to random chance.

\begin{center}
    \includegraphics[width=0.7\linewidth]{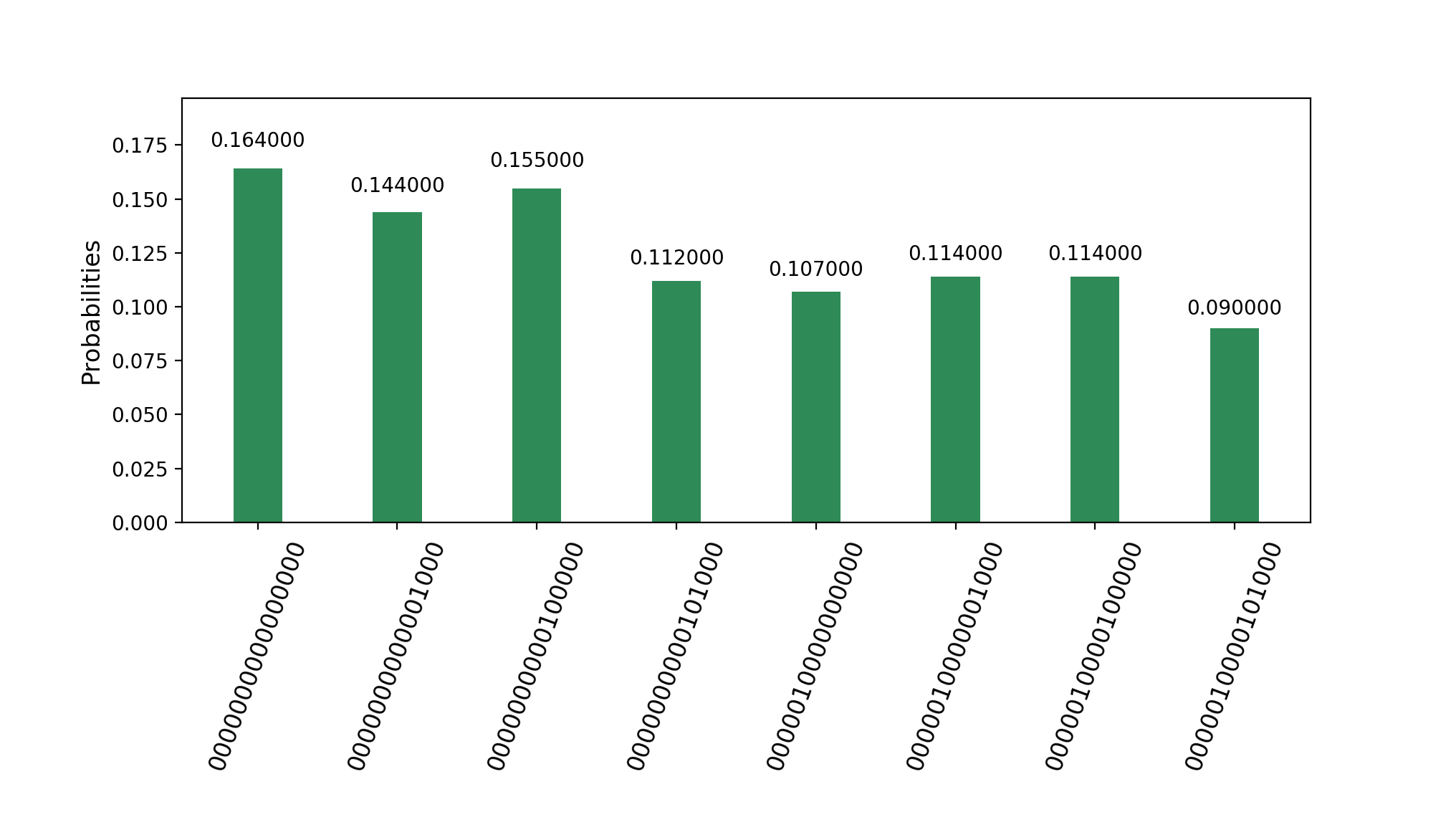}
    \captionsetup{width=0.8\textwidth}
    \captionof{figure}{An example of the histograms produced using QISKit's internal methods. While it does not provide much information regarding the single-qubit outputs, it allows for a visualisation of the lack of statistical variation between different output states (as given by the x-axis).}
\end{center}

Unfortunately, due to restrictions IBM places on the number of times an experiment can be run on its chip per day, each run could only be repeated two to three times, but again the standard deviation from the mean was less than 1\%, making it very likely that the magnitude of the errors would persist regardless of repetition. 

\subsection{Results Post-Optimization and Using Classical Processing of Measurements}

The results obtained as described above prompted a series of attempts at optimization as described in the \textbf{Methods} section. The results appear promising, as seen in Figure \ref{bestrun}, where despite the lack of improvement in qubits 1 and 3, the deviation of qubit 2 from the simulation values was reduced from 39.6\% to 25.2\%.

\begin{center}
    \includegraphics[width=0.7\linewidth]{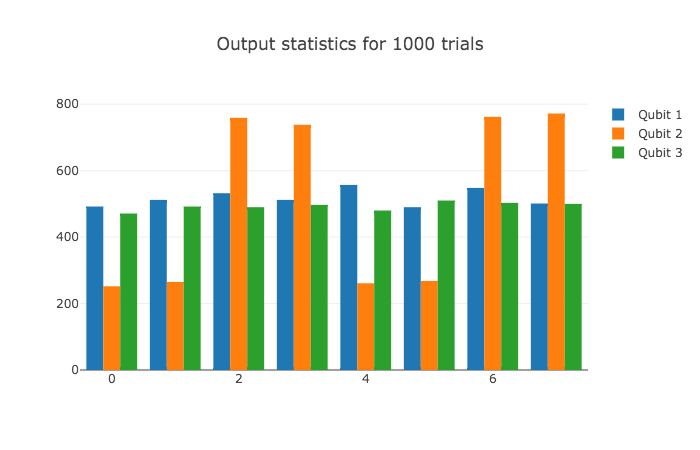}
    \captionsetup{width=0.8\textwidth}
    \captionof{figure}{A histogram of optimized IBM chip output qubit values for each classical input. The experiment was again run 1000 times for each input and the extracted values plotted using F\#.}
    \label{bestrun}
\end{center}

 This can easily be attributed to the fact that the second qubit has a far smaller number of  operations when calculating its corrected output value. It does not require to wait for Alice's qubit measurements or the extra gates that are needed to perform these (refer back to the original MBQC circuit diagram in Figure \ref{circuit}).
 
 \begin{table}[H]
\centering
\begin{tabu}{|p{0.5cm}|[2pt]p{0.5cm}|p{0.5cm}|p{0.5cm}|p{0.5cm}|p{0.5cm}|p{0.5cm}|p{0.5cm}|p{0.5cm}|}
\hline
\multirow{2}{*}{} & \multicolumn{8}{c|}{Classical input value} \\ \cline{2-9} 
                  & 0 &  1 & 2 & 3  & 4 & 5 & 6 & 7\\ \tabucline[2pt]{-}
Q1                  & 492 & 512 & 532 & 512 & 557 & 490 & 548 & 501 \\ \hline
Q2                  & 252 & 265 & 759 & 738 & 261 & 268 & 762 & 772\\ \hline
Q3                  & 471 & 492 & 490 & 497 & 480 & 510 & 503 & 500 \\ \hline
\end{tabu}
\caption{Table showing the number of times out of 1000 that a 1 was calculated as the value of output qubits in the optimized version of the IBM circuit. While the second qubit shows some statistical improvement, the computation as a whole does not provide useful results.}
\label{bestresulttable}
\end{table}

 It is difficult to determine how much of the final error was due to decoherence over time and how much of it was due to the number of operations applied to each qubit. It is likely that the final circuit runs, as optimized internally by QISKit, were the best possible version of results that could be produced. An attempt to add barriers, which act by confining operations to a particular order of execution only made the results for the second qubit worse, meaning the small advantage of performing some operations on qubits 1 and 3 sooner came at the cost of the coherence of the $2^{nd}$ with no tangible gain for the coherence of the other two. Further attempts at optimization by moving the measurements of the client to be before the other measurements and similar methods all seem to produce the same results, so nothing else could be done to optimize the results.
 
\section{Conclusions and Future Research}

The data obtained during the experimental runs on the IBM chip and the classical simulations lends itself to several different conclusions, both concerning the QFHE protocol and the future of quantum computation as a commercial service. One positive result is the success of classical simulations. While this does not reflect the difficulty of implementing it in real quantum hardware architectures, it does at least show that the scheme can perform as intended, in theory, on simple computations. There is little reason to doubt that the transfer of this functionality to real technology is possible, if the technology is to perform as predicted. The problems concerning this, however, merit some discussion. 

IBM's current quantum technology does not seem to provide enough fault-tolerance to test the QFHE protocol. The issue of decoherence and the errors induced in computations through performing operations such as gates and measurements will likely be one of the greatest challenges in the future of quantum computation. The simulations through classical means, both using LIQ$Ui\ket{}$ and the QISKit simulator exist to show that, within the confines of quantum technology performing as it should with 100\% fault-tolerance, the protocol should work as intended. The fact that the real chip does not confirm this can only be explained as either quantum operations in the hardware not being equivalent to those in theory, as used by the simulators, or through the fragility of quantum systems and the high level of errors induced in these computations without the resources to error-correct. The former, while not impossible, would still imply that something about these technologies is inherently different to what the theory predicts and makes less sense when considering that smaller examples of code using the same technology and the same theoretical premises appear to produce correct results as in example circuits provided by the QISKit software package\footnote{\url{https://github.com/QISKit/qiskit-sdk-py/tree/master/examples/python}[Last accessed: 26/11/2017]}. However this is part of a larger problem of quantum verification which is too vast to be discussed here (a recent review paper on the subject gives a good overview of the current state of quantum verification research: \cite{verification}). If increasing the scale of the computation begins to make it fail, it's far more likely to be a growth in error than the theoretical premises failing, partcularly for examples of this size. Presuming this allows for a discussion about the cause of these errors and how to better simulate and finally apply the QFHE protocol once the technological resources become available.

The obvious solution, beyond improving the fault-tolerance of the quantum hardware itself, is to implement error-correcting codes. Shor's code in particular is promising- it can correct any corrupting channel as long as only one qubit is affected. However, it also requires eight additional qubits for one qubit of useful information\cite{nielsen2000quantum}, so IBM's current 16-qubit chip does not have nearly enough in terms of resources to involve this type of correction and perform a useful computation, although once it does, the same example circuit could be run to check whether there is any development. 

The second possibility of improvement is that of avoiding decoherence by extracting information from qubits as quickly as possible. While remaining errors are unavoidable once the number of operations in the circuit has been optimized, giving it less time to lose information through contact with the environment is a promising option. The most likely reason that the first and third output qubits give the worst results is because of the time it takes between the initiation of the Bell Pairs and their actual measurement- as well as the number of gates the conditional measurement requires. In the case of QFHE, the client would supposedly be able to measure their qubit immediately after the server has measured their half and sent it back, reducing the amount of time for decoherence (if the transmission of the qubit does not make the loss of information even worse than in the current experiment). 

Furthermore, the general idea of QFHE as well as other MBQC-based protocols is that the qubits could be entangled and measured layer-by-layer along the direction of the flow. This means that the time between initializing and measuring the qubit could be made as short as the span of two entanglements and 2-3 gate operations. This is, in fact, possible with the current hardware, but since the computation is not actually networked, Alice and Bob cannot be performing measurement simultaneously, leaving the measurement of the Bell pair qubit to accumulate decoherence-induced errors. 

An interesting possibility to consider, once the hardware for it has been developed, is the reusal of qubits in MBQC computations after they'd been measured. If a true networked computation were a possibility, then depending on the dimensions of the computation, each layer of the MBQC graph could be re-initialised once the qubits in it have been measured and the values stored or transmitted. This would allow for infinite depth in a computation with no marginal increase in decoherence of each qubit. 

Finally, although this has already been alluded to above, it is important to discuss how the experiments run on the IBM chip do not reflect the actual implementation of QFHE in a number of important ways. The classical simulations are a better reflection bar the errors and decoherence, but the actual hardware results are for a single circuit that has been transformed in so many ways that the original scheme is barely recognizable. The fact that the client cannot measure their Bell pair half is important not just in terms of reducing the decoherence time, but also in predicting what kind of impact the transmission of the qubit to her will have. If the errors in gate operations and measurements are this large, should future quantum hardware be adapted to reduce the errors of qubit transmission and reception as well? There is promising technology already being developed, such as by researchers in DELFT, who have managed to entangle two electrons separated by a distance of 3 meters using spin-photons\cite{Bernien2013}. Perhaps these sort of quantum networks will be a far better place to test and implement security protocols based in MBQC.

The development of quantum information hardware is currently moving at an unprecedented pace, with IBM set to release a 20-qubit chip by the end of the year and a 50-qubit device soon after\cite{ibm20qubits}. Despite the shortcomings apparent in current technologies, it looks to be a bright future for companies like IBM and Google that continue to invest in making their devices more fault-tolerant and adding functionality. Once the quantum resources exist, there is every reason to expect that protocols like UBQC and QFHE will be used extensively in areas ranging from quantum encryption to the development and understanding of new pharmaceuticals and simulations of molecules.

\section{Acknowledgements}

I would like to thank Elham Kashefi for giving me the chance to investigate the most fascinating, cutting-edge technologies ever developed as well as Petros Wallden for all the help in understanding his protocol as well as being a great professor. Most of all, thank you to Andru Gheorghiu without whom I would have barely grazed the surface both in understanding the concepts for the project and in actually being able to execute it. Like a padawan following the jedi master, I can only hope to continue the legacy.
 
\printbibliography

\newpage
\appendix
\section*{Appendices}
\addcontentsline{toc}{section}{Appendices}
\renewcommand{\thesubsection}{\Alph{subsection}}

\subsection{Samples of Code}
\subsubsection{QISKit Code for IBM}

\lstset{
    language=Python,
    basicstyle=\scriptsize,
    frame=single,
    breaklines=true,
    caption={Example of extracting relevant data from the IBM chip output and processing it to get corrected single-qubit outputs in the final, optimized version of the circuit. The states and their frequency are initially given as string-integer pairs in a Dictionary object.}
}
\begin{lstlisting}

# RUN THE EXPERIMENT
result = qp.execute(['MBQC'], backend='ibmqx5', shots = 1000, timeout = 600)

# get the data obtained in the experiment
counts = result.get_data('MBQC') 
#ran_qasm = result.get_ran_qasm('MBQC')
#print(ran_qasm)

# Extract relevant counts for qubit outcomes from the data obtained
counts = counts['counts']
counts = counts.items()
for i in counts:
    val = i[0] #state measured
    times = i[1] #times it was measured
    if (times != 0):
        out0 = (ord(val[0])-48)
        out2 = (ord(val[2])-48)
        out3 = (ord(val[3])-48)
        out5 = (ord(val[5])-48)
        out6 = (ord(val[6])-48)
        out7 = (ord(val[7])-48)
        out8 = (ord(val[8])-48)
        out9 = (ord(val[9])-48)
        out10 = (ord(val[10])-48)
        out12 = (ord(val[12])-48)
        out13 = (ord(val[13])-48)
        out15 = (ord(val[15])-48)

        # get corrected final outcomes
        res1 += ((out15 + out0 + out12 + out13)%2)*times
        res2 += ((out8 + out7 + out6)%2)*times
        res3 += ((out2 + out3 + out10 + out9)%2)*times
\end{lstlisting}

\lstset{
    language=Python,
    basicstyle=\scriptsize,
    frame=single,
    breaklines=true,
    caption={Sample of optimized code for the IBM chip, wherein many Hadamard operations have been removed and subsequently entanglements have been de-constructed into CNOT and Hadamard operations.}
}
\begin{lstlisting}

# Hadamard everything
mbqc.h(q1[0])
mbqc.h(q1[1])
mbqc.h(q1[3])
mbqc.h(q1[4])
mbqc.h(q1[6])
mbqc.h(q1[8])
mbqc.h(q1[9])
mbqc.h(q1[12])
mbqc.h(q1[15])

# first Bell pair
mbqc.cx(q1[15], q1[2])
mbqc.x(q1[2])
#mbqc.barrier()

# second Bell pair
mbqc.cx(q1[12], q1[5])
mbqc.x(q1[5])
#mbqc.barrier()

# entangle qubits

mbqc.cz(q1[15], q1[0])
mbqc.cx(q1[8], q1[7])
mbqc.h(q1[7])
mbqc.cx(q1[12], q1[13])
mbqc.h(q1[13])

mbqc.cx(q1[15], q1[14])
mbqc.cx(q1[7], q1[10])
mbqc.cx(q1[12], q1[11])

mbqc.cx(q1[3], q1[14])
mbqc.cx(q1[9], q1[10])
mbqc.cx(q1[6], q1[11])

mbqc.h(q1[11])
mbqc.cx(q1[11], q1[10])
mbqc.h(q1[11])

\end{lstlisting}

\subsubsection{LIQ$Ui\ket{}$ Code}

\begin{center}
    \includegraphics[width=0.9\linewidth]{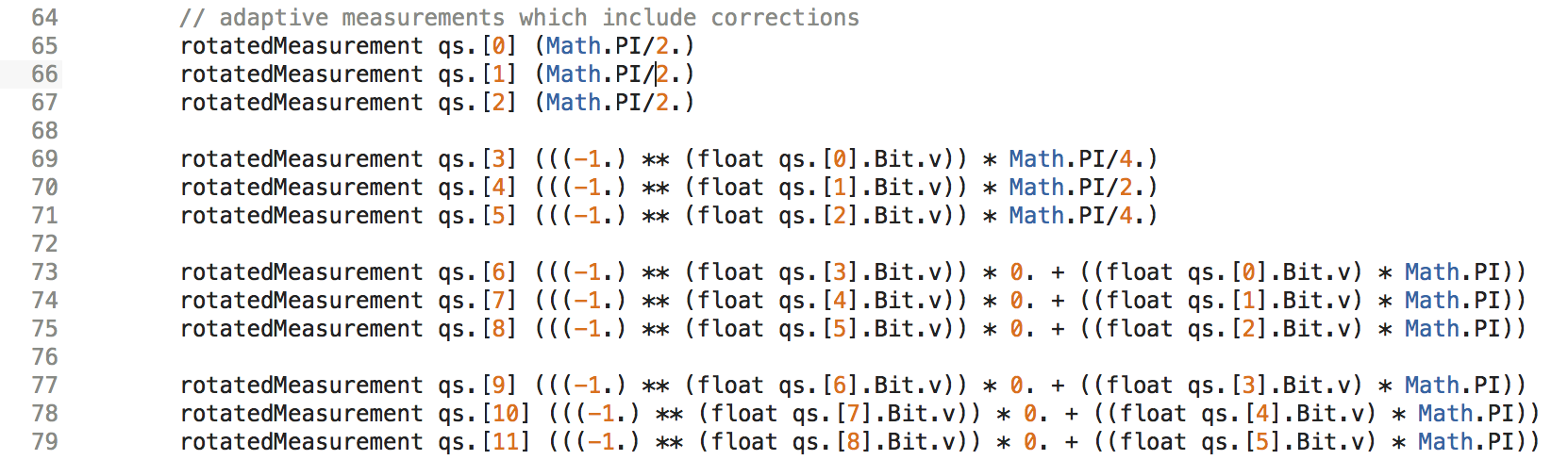}
    \captionof{figure}{Code for performing corrections interactively with consequent angles based on measurement outcomes of previous qubits. The method \texttt{rotatedMeasurement} was written specifically for the simulation and takes two arguments: the qubit to operate on and the angle at which the measurement operation should be performed.}
    \label{intercor}
\end{center}

\begin{center}
    \includegraphics[width=0.9\linewidth]{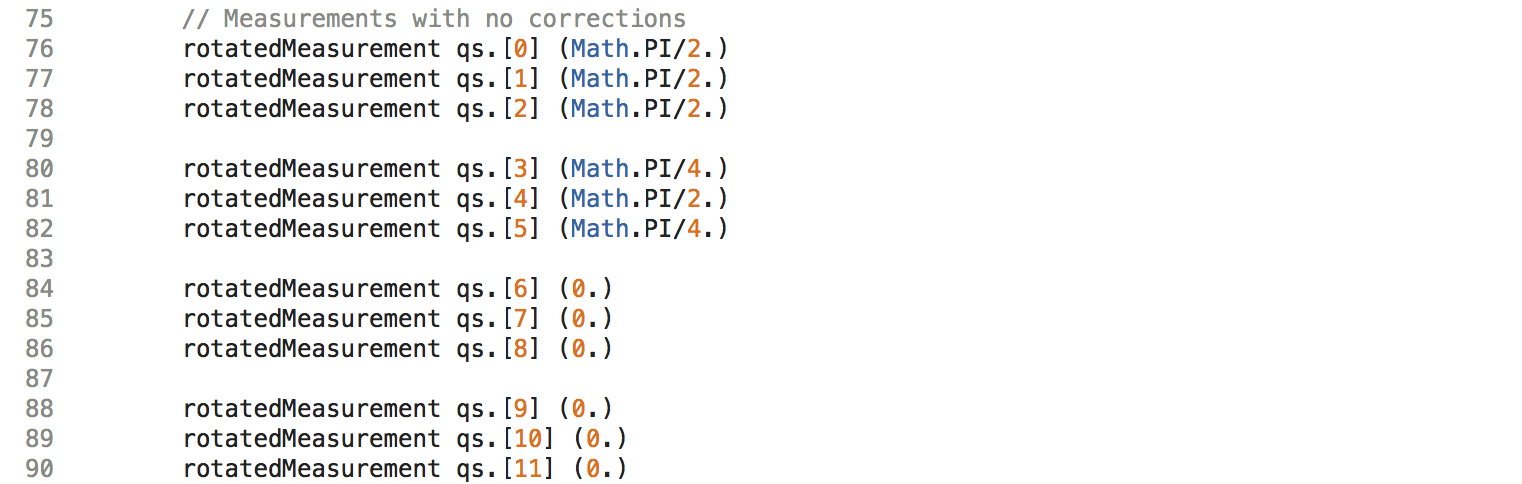}
    \captionof{figure}{Code of the measurements of the same qubits at the default angles provided at the beginning of the computation without any interactive corrections.}
    \label{nocor}
\end{center}

\begin{center}
    \includegraphics[width=0.9\linewidth]{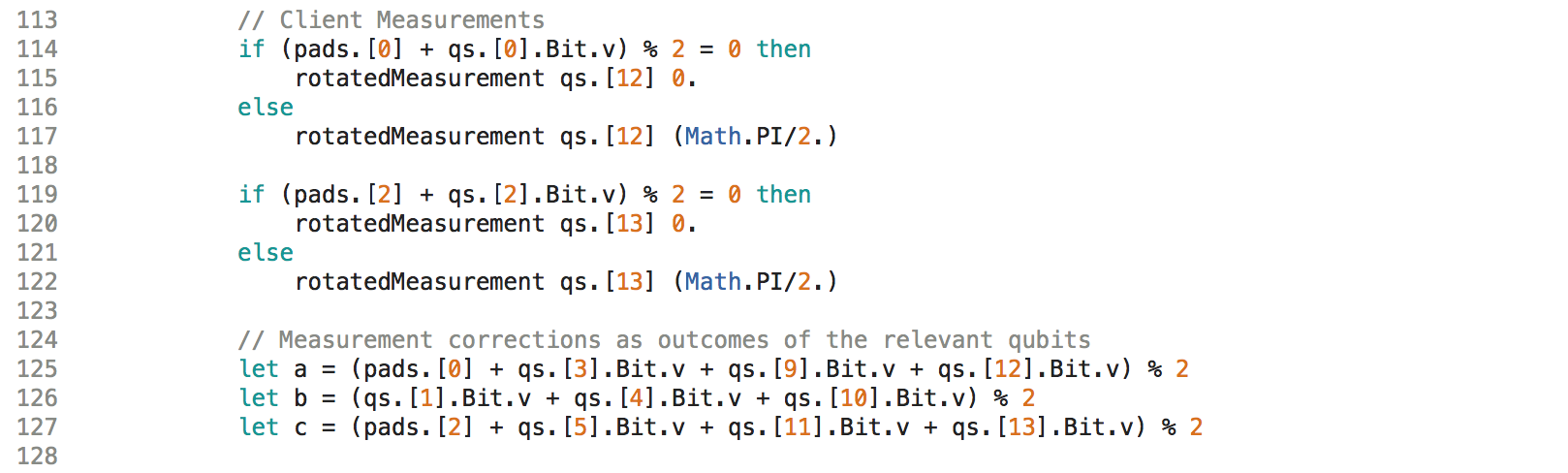}
    \captionof{figure}{Measurements of the Bell pair qubits performed by the client and the subsquent final amplitudes of the relevant qubits calculated with deferred corrections. Variables \texttt{a}, \texttt{b} and \texttt{c} are then the final corrected outcomes of output qubits 1, 2 and 3 respectively.}
    \label{client}
\end{center}

\newpage
\subsection{Common Quantum Gates}

\begin{center}
    \includegraphics[width=0.5\linewidth]{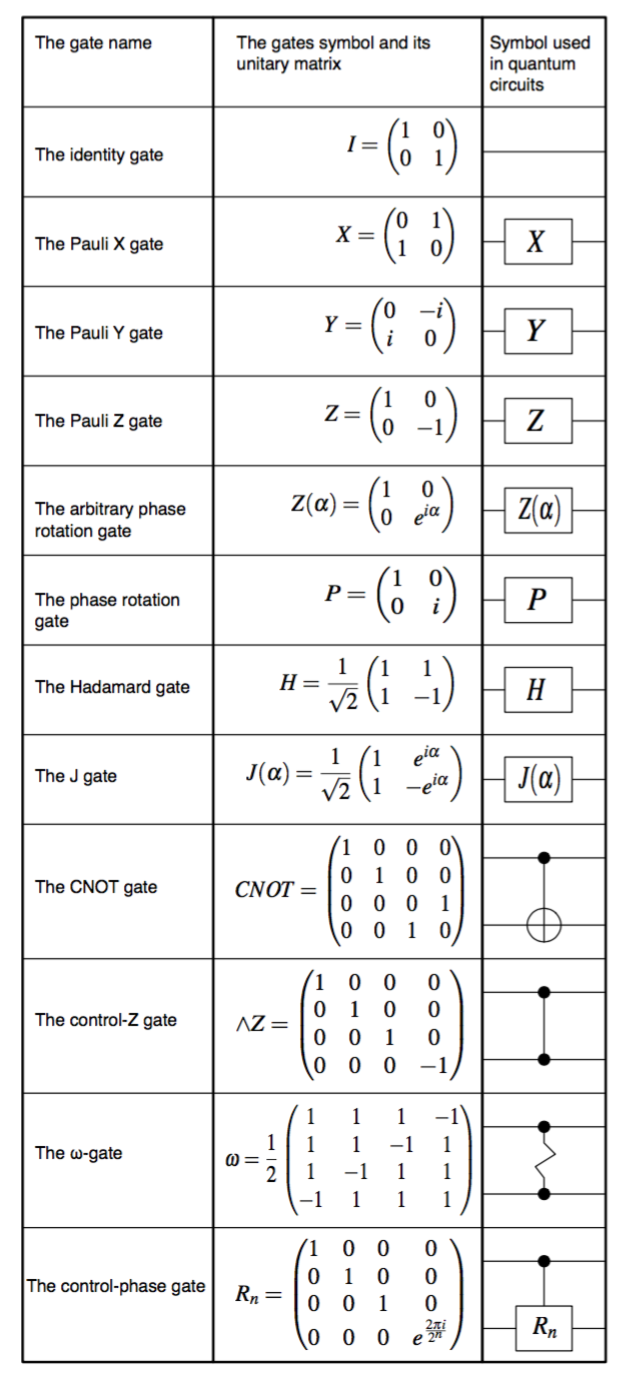}
    \captionof{figure}{Some of the most common gates used in quantum computation and their representations both as matrices and in quantum circuits.}
\end{center}

\end{document}